\documentclass[
reprint, superscriptaddress, amsmath,amssymb, aps, pra,
]{revtex4-2}

\usepackage{graphicx}
\usepackage{dcolumn}
\usepackage{bm}

\newcommand{\pare}[1]{\left( #1 \right)}

\newcommand{\cor}[1]{\left[ #1 \right]}
\newcommand{\llav}[1]{\left\lbrace #1 \right\rbrace}

\begin{document}

\preprint{APS/123-QED}

\title{High-density three-dimensional holography using rapid modulation of light }%

\author{Jorge-Alberto Peralta-\'Angeles}
\email{jperalta@ciencias.unam.mx}
 \affiliation{Instituto de Ciencias Nucleares, Universidad Nacional Aut\'onoma de M\'exico}

\author{Mingyuan Hong}
 \affiliation{Quantum Photonics Laboratory, Department of Physics and Astronomy, Louisiana State University}


\author{Mario A. Quiroz-Ju\'{a}rez}
\affiliation{Centro de F\'isica Aplicada y Tecnolog\'ia Avanzada, Universidad Nacional Aut\'onoma de M\'exico, Boulevard Juriquilla 3001, Juriquilla, 76230 Quer\'etaro, M\'exico}

\author{Omar S. Maga\~na-Loaiza}
 \email{maganaloaiza@lsu.edu}
 \affiliation{Quantum Photonics Laboratory, Department of Physics and Astronomy, Louisiana State University}

\author{Roberto de J. Le\'on-Montiel}%
 \email{roberto.leon@nucleares.unam.mx}
 \affiliation{Instituto de Ciencias Nucleares, Universidad Nacional Aut\'onoma de M\'exico}%

\date{\today}

\begin{abstract}
One of the most common methods for reconstructing three-dimensional (3D) images of real objects is digital holography. This technique relies on the use of electro-optical devices that modify the phase or amplitude of light fields in a controlled manner, the so-called spatial light modulators. 
However, given that holography typically requires coherent light sources, a common problem with three-dimensional projection is the crosstalk between the layers that make up the 3D object. This limits full-depth control and directly affects image quality. Interestingly, in the past few years, several methods have proven to be effective in breaking layer crosstalk by erasing the spatial coherence of light. A drawback of such solutions is that, in many cases, additional optical resources are required to achieve such a task. In this work, we present a method for high-density reconstruction of three-dimensional objects using rapid modulation of light fields by means of digital micromirror devices (DMDs). The 3D reconstruction is performed by discretizing the object into multiplane light-point contours, where the resolution of the contours is controlled by the density of the light points. This allows us to achieve point separations, in the transverse plane, as small as 100 $\mu$m. The high refresh rate of the DMD ($\sim 10$ kHz) allows for a reconstruction where each point of the 3D image is spatially and temporally controlled by independent amplitude holograms, thus effectively eliminating coherence-induced multiplane crosstalk without the need for additional optical elements. Because of its simplicity and versatility, we believe that our method provides a practical route toward compact, high-resolution 3D holographic projectors.
 
\end{abstract}

\maketitle

\section{Introduction}

Three-dimensional (3D) holography has emerged as a powerful technique for reconstructing volumetric objects using light \cite{Pan:16}. Traditional approaches rely on holographic plates or charge-coupled device (CCD) sensors to record the interference pattern of coherent light scattered by an object. Unfortunately, these methods can be slow and impractical for dynamic or high-resolution applications \cite{Rhisheekesan:24}. The advent of digital holography has greatly improved this process by enabling faster acquisition, digital storage, and flexible manipulation of the phase and amplitude of light \cite{Rhisheekesan:24, Liu:16, Ayoub:21}. These advances have expanded the potential of holography in fields such as biomedical imaging, industrial inspection, and immersive display technologies \cite{Yue:13, Chonglei:21, Sando:18, Ouyang:23, Xiong:21, Jang:24, Wang:24}.

A key component of digital holography is the use of spatial light modulators (SLMs), among which digital micromirror devices (DMDs) have attracted increasing attention due to their fast refresh rates \cite{Mirhosseini:13, Perez-Garcia:2022}. DMDs have enabled the rapid modulation of amplitude and phase of optical beams \cite{Mirhosseini:13}. This possibility has enabled the engineering of structured light fields, including beams with orbital angular momentum and complex spatial patterns, as demonstrated in quantum optics, imaging, and beam shaping applications \cite{Mirhosseini:13, Lerner:12, Perez-Garcia:2022}. Recent studies have further highlighted the versatility of DMDs in quantum and statistical imaging tasks. For example, quantum statistical imaging has been shown to achieve super-resolution by analyzing spatial photon correlations sampled through programmable DMD patterns \cite{Bhusal:22}. Subsequent work demonstrated the ability to engineer spatial light modes with tunable photon-number statistics using random phase modulation on a DMD \cite{Hong:23}. Most recently, DMD-generated random binary patterns have been used to encode thermal light fields, enabling the extraction of quantum images from noisy natural light sources \cite{Mostafavi:25}. These developments demonstrate the potential of DMDs not only as display elements but also as tools for high-speed spatial light control across diverse imaging techniques \cite{Dudley:03, Zhu:12}.

DMDs capabilities can be successfully used to display 3D scenes \cite{Huebschman:03, Yu:23}; however, to optically reconstruct a good quality image of a 3D object, it is fundamental to have full depth control \cite{Makey:19}. More specifically, to reconstruct a 3D object, multiple planes of the object must be successively projected, each with its respective contour or cross section. Given that holography typically requires coherent light sources, one of the principal drawbacks of 3D image reconstruction is the inter-plane crosstalk caused by the spatial coherence of light. This ultimately affects the quality of the reconstructed object \cite{Makey:19,Yu:23}. Different methods have been proposed to break the coherence-induced inter-plane crosstalk and achieve full depth control. These include using photon sieves \cite{Park:19} and scattering media \cite{Yu:23,Yu:17}. We remark that, although these methods provide an effective way to break the spatial coherence of light, they require additional elements that tend to make the experimental deployment of the holography protocol more complex. Other methods rely on adding random phases to the image planes \cite{Makey:19}, or on the projection of multiple images perpendicular to the hologram plane \cite{Dorrah:23}.

Here, we report a method for high-density three-dimensional holographic reconstruction using rapid modulation of light fields via a DMD. By discretizing a 3D object into multiplane light-point contours and encoding each light point with an independent amplitude hologram, we eliminate coherence-induced crosstalk and achieve fine spatial resolution with point separations as small as 100 $\mu$m. The high refresh rate of the DMD ($\sim$ 10 kHz) allows for dynamic, plane-by-plane projection of holograms without requiring additional optical components. Our experimental results confirm the Gaussian fidelity of the projected beams across multiple depths through spatial profile measurements and fitting. Remarkably, we also demonstrate 3D holographic reconstructions of multiple solid objects, including a pyramid, cone, and double cone, with progressively increasing contour densities. Our approach offers a compact and efficient route toward real-time, high-quality 3D holographic displays with enhanced depth control.

\section{Results and discussion }

\subsection{Holographic projection method}

Our holographic projection protocol [depicted in Fig. \ref{fig:ExpSetup}a] starts with the preparation of the 3D object that is to be reconstructed. We first discretize a solid 3D-rendered object into a set of planes that are perpendicular to the light propagation axis. In this arrangement, the $i$th plane (or slice) of the object is located at the position $z_i$ in the propagation axis. Likewise, the solid contours in each plane are reconstructed using dots that form a discretized contour. Note that the positions of the dots are fully characterized by the coordinates $(x_{i},y_{i},z_{i})$. Once this process is completed, the optical reconstruction of the object can be performed by directing light spots into specific positions that represent the object contours at different depths.

The spots used to perform the 3D reconstruction correspond to light beams bearing a Gaussian intensity profile in the transverse ($x$-$y$) plane. The waist at which the field amplitude falls to $1/e$ is given by $w_0$. Mathematically, each spot at a $z_{i}$ distance can be described by writing
\begin{equation}
 \psi_{z_{i}}(x,y) = \mathrm{exp}\left[-\frac{\pare{x-x_{i}}^2+\pare{y-y_{i}}^2}{w_0^2}\right],
 \label{eqn:gauss_beam}
\end{equation}
where we have assumed a unitary maximum amplitude for each beam. Note that light spots can be thought of as Gaussian beams being focused at the $(x_{i},y_{i})$ position by lenses with focal length equal to $z_{i}$. We exploit this idea in order to prepare each of the light spots by means of holograms displayed in the DMD.

To create the hologram of each spot, we assume that all light fields are described by Eq. (\ref{eqn:gauss_beam}). In the Fourier space, this expression takes the form 
\begin{eqnarray}
 \tilde{\psi}_{z_{i}}(f_x, f_y) & = &  \mathcal{F}\llav{\psi_{z_{i}}(x,y)}, \nonumber \\
 &=& \int_{-\infty}^{\infty}\int_{-\infty}^{\infty}\psi_{z_{i}}(x,y)e^{-i2\pi\pare{f_{x}x + f_{y}y}}dxdy, \nonumber \\
 &=& \pi w_{0}^{2}e^{-i2\pi\pare{f_{x}x_{i}+f_{y}y_{i}}}e^{-\pi^{2}w_{0}^{2}\pare{f_{x}^{2}+f_{y}^{2}}}.
 \label{eqn:gauss_beam_FT}
\end{eqnarray}
We then make use of the angular spectrum method \cite{Good05}, to back propagate the field to the origin $z=0$, which corresponds to the position of the DMD. The back propagation can be performed by applying the point spread function, in the Fresnel limit, given by \cite{Good05}
\begin{equation}
 H_{-z_{i}}(f_x,f_y) = \mathrm{exp}\left(-iz_{i}k\sqrt{1-\lambda^2f_x^2 - \lambda^2f_y^2}\right),
 \label{eqn:transfer_function}
\end{equation}
where $k=2\pi/\lambda$, and $\lambda$ is the wavelength of the light beam. The field at the DMD's position is thus obtained by inverse Fourier transforming the product of Eqs. (2) and (3), i.e.,  
\begin{equation}
 \psi_{0}(x,y) = \mathcal{F}^{-1}\llav{\tilde{\psi}_{z_{i}}(f_x, f_y)H_{-z_{i}}(f_x,f_y)}.
 \label{eqn:gauss_beam_FT_propagated}
\end{equation}
Finally, we create the hologram of the field by following the Kinoform method \cite{Good05}. To this end, we prepare the phase object $h(x,y)=\exp\cor{-i2\pi*\text{angle}\pare{\psi_{0}}}$, where $\text{angle}\pare{\cdots}$ stands for the phase of the field $\psi_{0}$. Upon binarization of $h(x,y)$ (see Ref. \cite{Mirhosseini:13} for details), the holograms for each of the light spots that make up the 3D object are ready to be deployed in the DMD. 






\subsection{Experimental setup}

\begin{figure*}
 \includegraphics[width=0.98\textwidth]{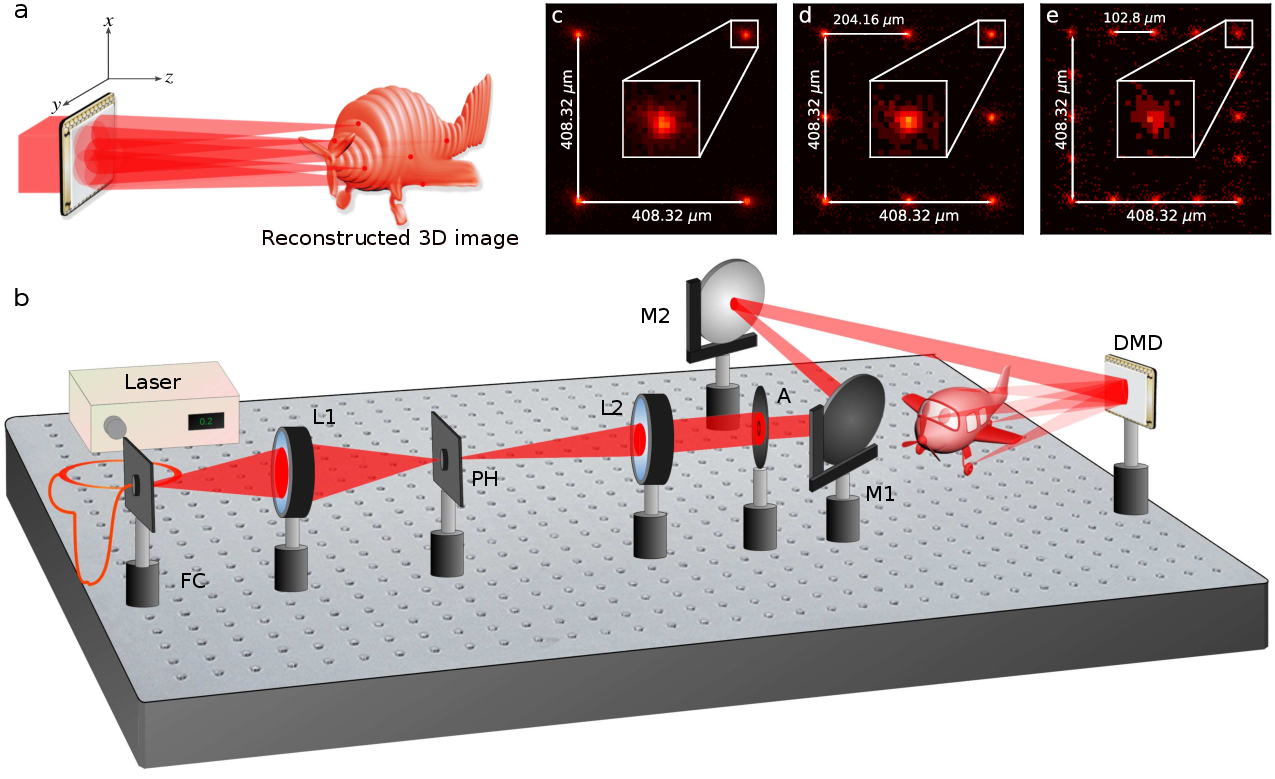}
 \caption{
 \textbf{Holographic projection system}. \textbf{a} Working principle: A 3D object is reconstructed by means of discretized contours at planes perpendicular to the reconstruction axis. \textbf{b}, Experimental setup. A DMD is illuminated with a laser beam through a spatial filter made up
 with two lenses (L1 and L2) and a pinhole (PH) to obtain a Gaussian beam. An aperture is used to set the beam size which is directed
 to the DMD with the mirrors M1 and M2. Finally, the planes or slices of the reconstructed 3D object are captured with a charge-coupled-device
 (CCD) camera not shown in diagram.
 \textbf{c, d, e}, squares with
 different light spots densities formed by lense holograms. All squares are coded to have a 400 $\mu$m by side and each spot
 is encoded to have a 25 $\mu$m waist. The first square is composed by four spots separated by 408.32 $\mu$m, the second one
 is prepared to have three spots by side, separated by 204.16 $\mu$m. The last one is encoded to have five squares by side, separated by 102.8 $\mu$m.
 }
 \label{fig:ExpSetup}
\end{figure*}

The setup for the implementation of our 3D holographic projection protocol is shown in Fig. \ref{fig:ExpSetup}b. As light source, we use a continuous-wave 637 nm fiber-coupled laser beam. A spatial filter---consisting of lenses L1 and L2 separated by a distance equal to the sum of their focal lengths and a pinhole (PH) located at the focal length of lens L1---is used to clean and expand the beam. This guarantees that light illuminates the entire cross section of the DMD. Note that the beam size can be further adjusted by an aperture (A), located after lens L2. The mirrors M1 and M2 direct the light to the center of the DMD
(DLP Lightcrafter 6500 from Texas Instruments) to ensure accurate illumination. Finally, the light reflected from the DMD is observed at an angle of $\theta$=12° with respect to the incident beam. 

To reconstruct a high-quality image of a 3D object, the holographic elements used during light projection must contain most of the object's information with full depth control. However, this is generally limited by the coherent crosstalk and depth interval of the image planes. We alleviate the first drawback by using independent holograms for each of the light dots that make up the 3D object. The second limitation depends completely on the spatial properties of the holographic device, in our case the DMD. More specifically, the depth interval is limited by the depth of field (DoF) of each plane, which is determined by the diffraction angle and pixel count of the pixelated spatial light modulator. In our case, the DMD has a size of 1920$\times$1080 pixels and a pixel pitch of $p=7.6$ $\mu$m, thus its DoF limit corresponds to \cite{Yu:23} $DoF_{lim}=8p^2/ \lambda$.
For a $\lambda=637$ nm laser, we find that our optical system has a $DoF_{lim}=725.4$ $\mu$m. Additionally, the shortest focal length of the effective lens that can focus light using a square region of $N\times N$ pixels of the DMD, the so-called Fresnel zonal plate, is given by $f_c=Np^2 / \lambda$.
This expression tells us that if we were to use the largest square hologram that our DMD can hold (i.e. a hologram of $N=1080$ pixels per side), the shortest focal length attained by the system would be $f_c=97.93$ mm. This implies that the planes used for the holographic reconstruction of 3D objects must be placed at distances greater than $97.93$ mm from the DMD if all its pixels are used. With this in mind, the plane positions for each of the object's slices are computed by making use of the DoF equation for a pixelated modulator, i.e. \cite{Yu:23, Makey:19},
\begin{equation}
 DoF_i=\frac{8\lambda}{D^2}z_i^2,
 \label{eqn:DoFi}
\end{equation}
where $D=Np$ is the length of each side of the hologram, and $z_i$ is the distance at which the beam is focused. To accurately determine the position of the object's planes, the overlap between two successive planes needs to be avoided. This is done by considering that for a given plane located at $z_i$, the consecutive plane must be located at least half of its corresponding DoF. Mathematically, this can be written as
\begin{equation}
 z_{i+1} = z_i + \frac{DoF_i}{2} + \frac{DoF_{i+1}}{2}. 
 \label{eqn:zi+1}
\end{equation}
By solving Eq. (\ref{eqn:zi+1}) for an initial position $z_i$, it is straightforward to obtain the position $z_{i+1}$ for all ($i$+1) planes.

Finally, to capture the 3D object images, we use a CCD camera (DCU223M) synchronized with a motorized stage (PT3-Z8), controlled by a K-cube (KDC101) servo motor controller.  
For all our reconstructed objects, we use square holograms of $651\times 651$ pixels. This sets a limit in the focal length of the effective lenses displayed on the DMD of $f_{c} \sim$59 mm. Given the arrangement and size of all optical elements in our setup, we set the starting plane of the reconstructed 3D object at $z_0=80$ mm from the DMD. The motorized stage has a travel of 24 mm, which limits the number of planes that we can access with the camera. By using Eq. (\ref{eqn:zi+1}), one can find that, within the travel distance of the motor, we can access nine planes of the reconstructed object. 

\subsection{Beam characterization}

To efficiently project our holographic 3D images, we make use of the pycrafter6500 library \cite{Pozzi:17}. This Python library allows us to load the information of all light spots, through their corresponding binarized holograms (see section ``Holographic projection'' method for details), that are needed to form the 3D image.
%
%
We first explore the capabilities of our system in terms of spot density. To this end, we define a simple geometric figure, namely a rectangular cuboid and adjust the spot density. Figures \ref{fig:ExpSetup}c to \ref{fig:ExpSetup}e show one of the transverse planes of the cuboid, with a side length of $408.32$ $\mu$m. Note that our technique is able to increase the spot density without affecting the beam shape, which is typically caused by coherent crosstalk. More specifically, in Fig. \ref{fig:ExpSetup}c the density is two points per side, while Fig. \ref{fig:ExpSetup}d shows a square with three points per side, where the separation between spots is 204.16 $\mu$m. Finally, Fig. \ref{fig:ExpSetup}e shows a square with a density of five points per side, which leads to a separation between spots of 102.08 $\mu$m. It is worth pointing out that because the power of the laser source is fixed during the density change, the power of individual beams becomes lower as the number of points is increased. This could be relieved by increasing the power of the laser as one increases the spot density. Finally, we remark that although the DMD's manual sets the maximum number of holograms that can be loaded to 512, the actual number that we can load, given the complexity of the holograms, is 396. This sets the limit of how many spots can be used in the 3D image discretization process. 

\begin{figure}
 \includegraphics[width=1\columnwidth]{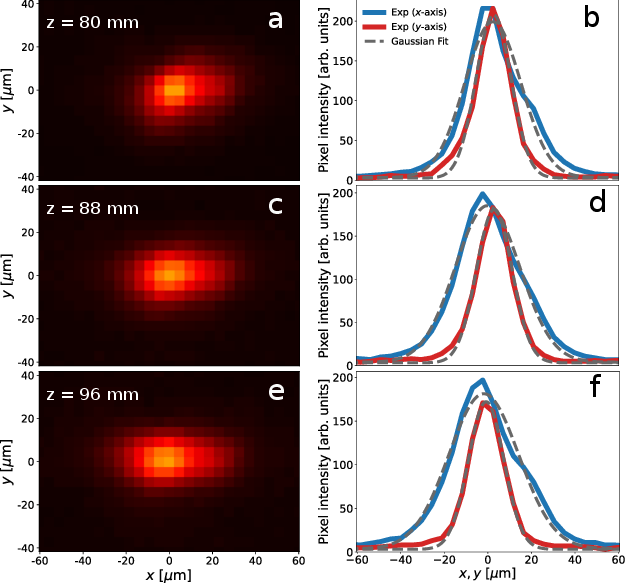}
 \caption{
  \textbf{Beam characterization}. Intensity distribution of light spots produced by the DMD at different planes: \textbf{a} $z=80$ mm, \textbf{c} $z=88$ mm, and \textbf{e} $z=96$ mm. \textbf{b}, \textbf{d} and \textbf{f} show the intensity profiles of the beams centered at the origin. The blue solid line depicts the beam along the x-axis centered at $y=0$, while the red solid line shows the shape of the beam along the y-axis centered at $x=0$. In all figures, the gray dashed lines correspond to Gaussian fittings. The R-squared values for each projection fitting, at different distances, are \textbf{b}: $R_{x}^{2}=0.97$, $R_{y}^{2}=0.99$; \textbf{d}: $R_{x}^{2}=0.99$, $R_{y}^{2}=0.99$; \textbf{f}: $R_{x}^{2}=0.98$, $R_{y}^{2}=0.99$. The beam images correspond to a single shot of a single hologram with a DMD exposure time of 150 $\mu$s and 0.2 mW laser power. 
  \label{fig:Spots}
 }
\end{figure}

In addition to spot density, we characterize the beam shape at different planes. For this, we load into the DMD holograms that produce Gaussian beams (with a waist $w_0=25$ $\mu$m) at three different distances, namely $z$ = 80, 88 and 96 mm. Figures \ref{fig:Spots}a, \ref{fig:Spots}c and \ref{fig:Spots}e show the recorded images. We note that the spots show an elongated intensity distribution along the x-axis, this can be attributed to variations in the DMD tilting around the $\pm$ 12° angle at which the micromirrors of the DMD rotate. As the propagation increases these angular variations become more dominant, resulting in greater beam elongation. We can quantify how close to a Gaussian the beams remain. For this we analyze their intensity distribution along the x- and y-axis. Figures \ref{fig:Spots}b, \ref{fig:Spots}d and \ref{fig:Spots}f show the intensity distributions of the beams centered at the origin. The blue solid line depicts the beam profile along the x-axis centered at $y=0$, while the red solid line shows the shape of the beam along the y-axis centered at $x=0$. In all figures, the gray dashed lines correspond to Gaussian fittings. The R-squared values for each projection fitting, at different distances, are \ref{fig:Spots}b: $R_{x}^{2}=0.97$, $R_{y}^{2}=0.99$; \ref{fig:Spots}d: $R_{x}^{2}=0.99$, $R_{y}^{2}=0.99$; \ref{fig:Spots}f: $R_{x}^{2}=0.98$ , $R_{y}^{2}=0.99$. These measurements show that, although not perfect, the beams remain Gaussian-like at different image planes.



\subsection{3D holographic reconstruction}

\begin{figure*}
\includegraphics[width=0.98\textwidth]{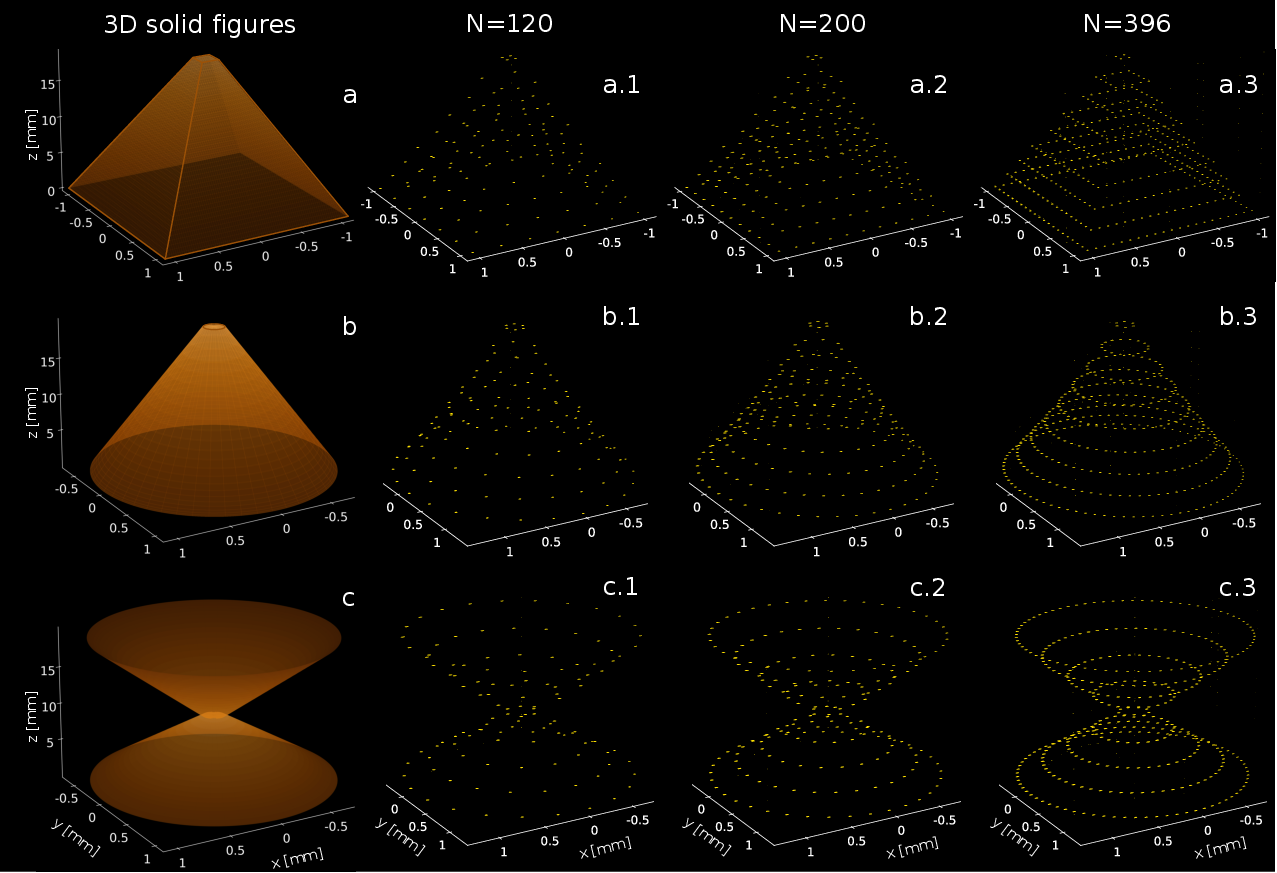}
\caption{\label{fig:3D_Figures} \textbf{Three-dimensional (3D) holographic reconstruction}. The first column shows the original 3D-rendered objects that are holographically reconstructed using the rapid modulation of light method. The shapes correspond to \textbf{a} solid pyramid, \textbf{b} solid cone, and \textbf{c} solid double-cone. The second, third and fourth columns show the holographic reconstruction of the 3D objects using 120, 200 and 396 light spots, respectively. The spot spacing for the low density object (\textbf{a}.1, \textbf{b}.1 and \textbf{c}.1) is $\sim$333 $\mu$m, while for the medium density (\textbf{a}.2, \textbf{b}.2 and \textbf{c}.2) the distance is $\sim$200 $\mu$m. Finally, for the high density reconstructions (\textbf{a}.3, \textbf{b}.3 and \textbf{c}.3) the distance between adjacent light spots is $\sim$100 $\mu$m. Note that our technique allows for increasing the density of light points of the holographic objects, without compromising the image quality with coherence-induced crosstalk effects.}
\end{figure*}

Our optical system allows for the holographic reconstruction of 3D objects. The discretization of the object is performed by carefully choosing the locations of the 396 light spots that can be loaded into the DMD. Note that the object is reconstructed in a rectangular volume of 24 mm (length) $\times$ 2 mm (width) $\times$ 2 mm (height). As described by Eq. (\ref{eqn:zi+1}), the 24 mm length (which corresponds to the maximum displacement of the camera) is divided into 9 imaging planes, while the width and height are limited by two parameters: (i) the number of pixels used for the spot holograms (651 x 651 pixels), and (ii) the region of the DMD where the holograms of far-from-the-origin light spots can be displayed.

We have prepared the 3D projection of three different objects, namely a pyramid, a cone and a double-cone. Figure \ref{fig:3D_Figures} shows the reconstruction of these three objects with increasingly higher spot density. The first column shows the 3D rendered solid figures, while the second, third and fourth columns show the holographic projections using 120, 200 and 396 light spots, respectively. The spot density is calculated taking into account the contour perimeter of each plane, the spacing between spots and the maximum number of correctly deployed holograms. The spot spacing for the low density (Figs. \ref{fig:3D_Figures} a.1, b.1 and c.1) is $\sim$333 $\mu$m so as to have 3 spots per mm, for the medium density (Figs. \ref{fig:3D_Figures} a.2, b.2 and c.2) the distance is $\sim$200 $\mu$m, which results in 5 spots per mm. Finally, for the high density (Figs. \ref{fig:3D_Figures} a.3, b.3 and c.3) the distance between adjacent points is $\sim$100 $\mu$m, thus resulting in 10 spots per mm. 

To properly capture the images, due to the fixed laser power (0.20 mW) and how the power is shared among all spots, the exposure time of the CCD camera was adjusted according to the exposure time of the holograms (105 $\mu$s) and the total number of holograms for the reconstructed object. For low density the exposure time was set to 25.2 ms, for medium density to 42 ms, and for high density 83.16 ms. In addition, to improve the sharpness of the images, we remark that the images shown in Fig. \ref{fig:3D_Figures} are obtained by averaging twenty shots for each plane. We conclude this section by pointing out how our technique allows for increasing the density of light points of the holographically reconstructed objects, without compromising the image quality. This is due to the lack of coherence-induced crosstalk between the light spots that produce the object.

\section*{\label{Conclusion} Conclusion}

In this work, we have presented a method for high-density reconstruction of three-dimensional objects using rapid modulation of light fields via DMDs. The 3D reconstruction is performed by discretizing the object into multiplane light-point contours, where the resolution of the contours is controlled by the density of light spots. By carefully balancing the high refresh rate of the DMD ($\sim 10$ kHz)---which corresponds to the inverse of its exposure time (105 $\mu$s)---and the number of holograms that can be loaded into the memory of the DMD's driver (396 holograms), we achieve point separations, in the transverse plane of the object, of up to 100 $\mu$m. Because the holographic proceeds by rapidly projecting independent holograms for each light spot, our technique intrinsically eliminates any coherence-induced crosstalk effect, thus allowing us to increase the density of light points without compromising the image quality. Because of its simplicity and versatility, we believe that our method provides a practical route towards compact, high-resolution 3D holographic projectors.

\section*{Acknowledgments}
This work was supported by Dirección General de Asuntos del Personal Académico (DGAPA) through the project UNAM-PAPIIT IN101623 and IA103325. M.H. and O.S.M.L. acknowledge support by the Office of
the Under Secretary of Defense for Research and Engineering, administered by the Air Force Office of Scientific Research, under award number FA9550-24-1-0226.
J.-A.P.-Á. is grateful for financial support through a postdoctoral fellowship from SECIHTI.
M.A.Q.-J. thankfully acknowledges financial support by SECIHTI under the Project CF-2023-I-1496.


\bibliography{biblio}

\newpage
\appendix

\end{document}